\documentclass[twocolumn,showpacs,preprintnumbers,amsmath,amssymb]{revtex4}
\usepackage{graphicx}

\newcommand{\gtsim}{\protect\raisebox{-0.5ex}{$\:\stackrel{\textstyle >}{\sim}\:$}}
\begin{document}
\title{Electron-phonon coupling and spin-charge separation in one-dimensional Mott insulators}
\author{H. Matsueda${}^{a}$}
\altaffiliation[Present adress: ]{Department of Physics, Tohoku University, Sendai 980-8578, Japan}
\email{matsueda@cmpt.phys.tohoku.ac.jp}
\author{T. Tohyama${}^{a}$}
\author{S. Maekawa${}^{a,b}$}
\affiliation{
${}^{a}$Institute for Materials Research, Tohoku University, Sendai 980-8577, Japan, \\
${}^{b}$CREST, Japan Science and Technology Agency (JST), Kawaguchi 332-0012, Japan}
\date{\today}
\begin{abstract}
We examine the single-particle excitation spectrum in the one-dimensional Hubbard-Holstein model at half-filling by performing the dynamical density matrix renormalization group (DDMRG) calculation. The DDMRG results are interpreted as superposition of spectra for a spinless carrier dressed with phonons. The superposition is a consequence of robustness of the spin-charge separation against electron-phonon coupling. The separation is in contrast to the coupling between phonon and spin degrees of freedom in two-dimensional systems. We discuss implication of the results of the recent angle-resolved photoemission spectroscopy measurements on SrCuO${}_{2}$.
\end{abstract}
\maketitle

The interplay between electron correlation and electron-phonon coupling is one of the hot topics in the field of strongly correlated electron systems such as high-$T_{c}$ cuprates. Particularly in an electron-removal process from the Mott insulators, the electron-phonon coupling occurs due to charge imbalance around a created hole. Thus, the angle-resolved photoemission spectroscopy (ARPES) is a direct tool, and observes a quasiparticle dressed with phonons~\cite{Damascelli}. For the two-dimensional (2D) insulating cuprates, the ARPES experiments have revealed a broad low-energy peak that is interpleted as a result of disappearance of the quasiparticle weight due to the coupling~\cite{Rosch,Shen,Mishchenko}.

The quasiparticle in the 2D systems is not only dressed with phonon cloud, but also dressed with antiferromagnetic (AF) spin fluctuation. Thus, the effect of phonon on the spectrum is affected by the spin configuration of the background. However, in one-dimensional (1D) correlated electron systems, a photohole created by ARPES decays into spinon and holon due to the spin-charge separation~\cite{Kim2}. Therefore, the effect of phonon on the spectrum strongly depends on whether the spin-charge separation is robust against the electron-phonon coupling. In this Letter, we examine the effect of phonon on the single-particle excitation spectrum in 1D Mott insulators.

Theoretically, the Hubbard-Holstein model is a basic model to study the interplay between electron correlation and electron-phonon coupling in cuprates~\cite{Khaliullin,Gunnarsson}. However, we have only limited information on the single-particle excitation spectrum in this model~\cite{Mishchenko,Tsutsui,Capone,Hohenadler2,Fehske,Bauml,Koller}. This is because it is hard to treat the infinite phononic degrees of freedom and electron correlation on an equal footing. In order to overcome the difficulty, we apply the dynamical density matrix renormalization group (DDMRG) method to the calculation of the spectra in the half-filled systems.

We find the following four characteristic features in the single-particle excitation spectrum: (i) a dip at high-binding energy side of the spinon branch, (ii) broad holon branch, (iii) decrease of the spectral weight of the spinon branch, and (iv) slight enhancement of the weight at low-binding energy side of the spinon branch. In the light of the spin-charge separation, these results are reproduced well by performing superposition of the spectra for a spinless carrier dressed with phonons. We examine the difference between the phonon effects on 1D and 2D systems. Implication of the results for the recent ARPES measurements on a 1D Mott insulator SrCuO${}_{2}$ is discussed.

The Hubbard-Holstein Hamiltonian in 1D is defined by
\begin{eqnarray}
H&=&-t\sum_{i,\sigma}( c^{\dagger}_{i,\sigma}c_{i+1,\sigma}+{\rm H.c.} ) \nonumber \\
&&+U\sum_{i}\left(n_{i,\uparrow}-\frac{1}{2}\right)\left(n_{i,\downarrow}-\frac{1}{2}\right) \nonumber \\
&&+\omega_{0}\sum_{i}b^{+}_{i}b_{i}-g\sum_{i}(b^{+}_{i}+b_{i})(n_{i}-1) , \label{HH}
\end{eqnarray}
where $c_{i,\sigma}^{\dagger}$ ($c_{i,\sigma}$) is the creation (annihilation) operator for an electron with spin $\sigma$ at site $i$, $b^{+}_{i}$ ($b_{i}$) is the creation (annihilation) operator for an Einstein phonon at site $i$, $n_{i}=n_{i,\uparrow}+n_{i,\downarrow}$, $n_{i,\sigma}=c_{i,\sigma}^{\dagger}c_{i,\sigma}$, $t$ is the hopping integral, $U$ is the on-site Coulomb repulsion, $\omega_{0}$ is the phonon frequency, and $g$ is the electron-phonon coupling constant. We take $U$ to be $U=8t$, which is an appropriate value for the cuprates. Here, we briefly mention how to construct the model Hamiltonian in the case of the cuprates. The lower Hubbard band of Eq. (\ref{HH}) corresponds to the Zhang-Rice singlet which is derived from a three-band model for Cu $3d_{x^{2}-y^{2}}$ and O $2p$ orbitals with lattice distortion. The distortion leads to change of the hopping integral of an electron between neighboring Cu and O orbitals. Due to the modulation of the hopping integral, the diagonal (Holstein) term dominates in the effective Hamiltonian for the Zhang-Rice singlet in comparison with off-diagonal terms~\cite{Khaliullin,Gunnarsson}.

We examine the single-particle excitation spectrum at zero temperature defined by
\begin{eqnarray}
A(k,\omega)=-\frac{1}{\pi}{\rm Im}\left< 0\left| c^{+}_{k,\uparrow}\frac{1}{E_{0}-\omega-H+i\gamma}c_{k,\uparrow}\right| 0\right> ,
\end{eqnarray}
where $c_{k,\uparrow}$ is the momentum representation of the electron operator $c_{i,\uparrow}$, $\left| 0\right>$ denotes the ground state with energy $E_{0}$, and $\gamma$ is a small positive number, which is taken to be $\gamma=0.1t$ in the present calculation.

Here, $A(k,\omega)$ is calculated by the finite-system DDMRG algorithm~\cite{White,Hallberg,Kuhner,Jeckelmann,Benthien,Matsueda,Shibata}. The system size $L$ is taken to be 20 lattice sites. The DMRG bases are truncated up to $m=400$ states from the density matrix for a mixed state of the ground state $\left|0\right>$, the final state after the one electron-removal process $c_{k,\uparrow}\left|0\right>$, and two correction vectors $(E_{0}-\omega-H+i\gamma)^{-1}c_{k,\uparrow}\left|0\right>$ for $\omega=\omega_{1}, \omega_{2}$, and $\omega_{2}-\omega_{1}=2\gamma$. It is technically useful to note that the convergent results are obtained when $A(k,\omega)$ and $A(k,\omega+2\gamma)$ are smoothly connected for a given $m$.

In order to keep numerical precision and reduce boudary effects, an open boundary condition is taken with potentials $-tn_{i}$ at the edges. In such a system, the momentum $k$ is defined by $k=n\pi/(L+1)$ with $n=1,2,...,L$. The momentum representation of $c_{l,\uparrow}$ is given by $c_{k,\uparrow}=\sqrt{2/(L+1)}\sum_{l}\sin(kl)c_{l,\uparrow}$~\cite{Benthien,Matsueda}. It is noted that the spectrum for $L=20$ presented here is similar to that for $L=120$ without the edge potential~\cite{Matsueda}.

In the DMRG calculation, electronic and phononic degrees of freedom at each site are devided into two different pseudo-sites. Furthermore, the phonon site is decomposed into a set of $N$ hard-core bosons, where the phonon states at each site are truncated up to $M=2^{N}$. The exact transformation between the ordinary boson operator and the set of the hard-core boson operators has been discussed in ref~\cite{Jeckelmann2}. After the transformation, the highest order term of the bosons is renormalized at first in some processes. The processes worsen numerical precision. Thus, we set up superblocks so that the final sweep process renormalizes the bosons in turn. Here, the maximum number of $N$ which we take is 4. The superblock is finally composed of $(1+N)L=100$ pseudo-sites.

\begin{figure}
\begin{center}
\includegraphics[width=13cm]{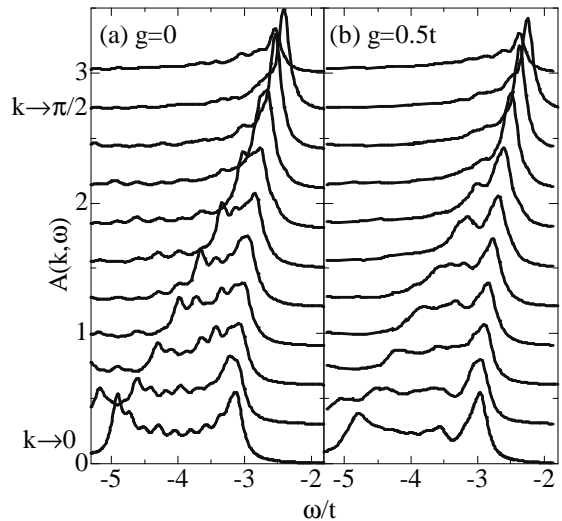}
\end{center}
\caption{$A(k,\omega)$ for the 1D Holstein-Hubbard model at half-filling. The momentum is taken from $\pi/21$ to $11\pi/21$.}
\end{figure}

\begin{figure}
\begin{center}
\includegraphics[width=13cm]{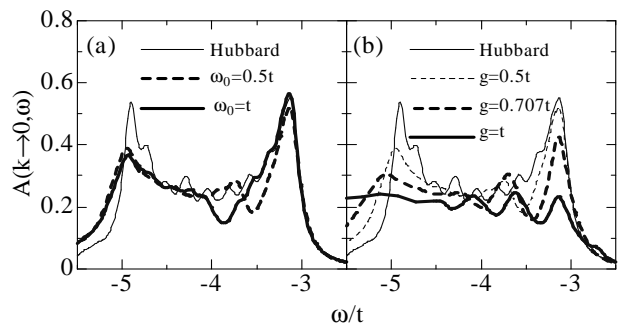}
\end{center}
\caption{(a) $A(k\rightarrow 0,\omega)$ for various $\omega_{0}$ values. For $\omega_{0}\ne 0$, $g$ is taken to be $0.5t$. (b) $A(k\rightarrow 0,\omega)$ with $\omega_{0}=0.5t$ for various $g$ values. For (a) and (b), the spectra with $g\ne 0$ are shifted so that the energy of the spinon branch is taken to be equal.}
\end{figure}

Figure 1 shows $A(k,\omega)$ with and without the electron-phonon couping $g$. The phonon frequency is taken to be $\omega_{0}=0.5t>\gamma =0.1t$ in order to see the effect of phonon clearly. However, the frequency for the cuprates is $\omega_{0}\gtsim \gamma$. Thus, we have confirmed that our conclusion does not change for $\omega_{0}=0.2t$. The origin of energy is located at the center of the Mott gap. In Fig. 1(a), two branches disperse, merging toward $k\rightarrow\pi/2$. These energy positions are $\omega/t=-3.13$ and $-4.90$ at $k=\pi/21$, which is the minimum momentum in the calculation. The branch located in low (high) binding energy side is deduced to be the spinon (holon) branch. Here, fine structures inside the branches come from the finite size effect. It is noted that the band width of the holon branch from $k=\pi/21$ to $k=10\pi/21$ depends on the open boundary condition and $U$. In Fig. 1(b), we find that the holon branch becomes broad due to the electron-phonon couping, while the spinon branch is sharp. We also find a 'peak-dip-hump' structure at the high-binding energy side of the spinon branch. The dip disperses like the spinon branch. As shown in Fig. 2(a), the energy of the dip (hump) position decreases with increasing $\omega_{0}$. The energy difference between the dip (hump) and the spinon branch is estimated to be $\omega_{0}$, which means that the peak-dip-hump structure is due to the phonon effect. Therefore, the orign of the peak-dip-hump structure provides information on the effect of phonon.

As $g$ goes from $0.5t$ to larger values with keeping $\omega_{0}=0.5t$, the spinon branch starts to broaden~\cite{Capone}. In Fig. 2(b), we show $A(k\rightarrow 0,\omega)$ for various $g$ values. The weight of the spinon branch decreases linearly as a function of $g$ for $g\ge\omega_{0}$. For $g=t$, the holon branch is completely smeared out. The peak-dip-hump structure observed in Fig. 1(b) develops into multipeaks whose positions are $\omega/t=-3.13$, $-3.61\sim -3.13-\omega_{0}$, and $-4.09\sim -3.13-2\omega_{0}$, respectively. Furthermore, a slight enhancement of the low-binding energy side of the spinon branch is seen. The energy difference between the enhancement and the band top ($k=\pi/2$) for $g=0$ is estimated to be $\omega_{0}$. Therefore, the three main peaks and the enhancement are also due to the phonon effect.

As mentioned in the previous two paragraphs, the following phonon effects appear on $A(k,\omega)$: (i) the peak-dip-hump structure, (ii) the broad holon branch, (iii) the decrease of the spectral weight of the spinon branch, and (iv) the enhancement of the low-binding energy side of the spinon branch. In particular, the characteristic energy scale of (i) and (iv) is $\omega_{0}$. In the light of the spin-charge separation, they can be interpreted as follows

\begin{figure}
\begin{center}
\includegraphics[width=13cm]{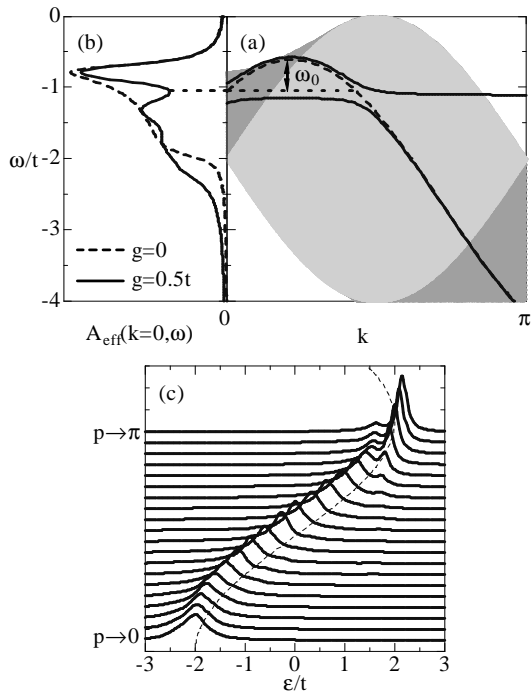}
\caption{(a) Shadded area: schematic view of the dispersion in a 1D Mott insulator; dashed (solid) line: one holon dispersion for $g=0$ ($g\ne 0$). A dotted line guides the eye. (b) $A_{\rm eff}(k=0,\omega)$; dashed (solid) line: $g=0$ ($g=\omega_{0}=0.5t$). Here, the broadening $\gamma$ is assumed. (c) $A_{h}(p-\pi,\varepsilon)$ for the 1D Holstein model, i.e. a model for a spinless carrier with Einstein phonons ($g=\omega_{0}=0.5t$). A dashed line is a cosine band with the width $4t$.}
\end{center}
\end{figure}

First, we consider the origins of (i) and (ii) shown in Fig. 1(b) for $g=0.5t=\omega_{0}$. Let us start with the dispersion in the Hubbard model illustrated in Fig. 3(a). According to the Bethe anzats solution, the dispersion is constructed by supersposition of a set of holon dispersions forming a cosine band with the width of $4t$~\cite{Sorella}. The superposition is a consequence of the spin-charge separation, because each of the holon dispersions is characterized by one spinon momentum. Therefore, an effective model of $A(k,\omega)$, $A_{\rm eff}(k,\omega)$, is constructed by putting the spectral weight for a spinless fermion, $A_{h}(p,\varepsilon)=\delta(\varepsilon-2t\cos p)$, on each of the holon dispersions. Since a top of the consine band is running along the spinon dispersion $\varepsilon_{s}(q+\pi/2)=-(\pi J/2)|\sin (q+\pi/2)|$ for $-\pi/2\le q\le\pi/2$, $A_{\rm eff}(k,\omega)$ is defined by
\begin{eqnarray}
A_{\rm eff}(k,\omega)=\sum_{q=-\pi/2}^{\pi/2}A_{h}\left(k-q,\omega+2t+\varepsilon_{s}\left(q+\frac{\pi}{2}\right)\right), \label{effective}
\end{eqnarray}
except for constant energy shift. A dashed line in Fig. 3(b) shows $A_{\rm eff}(k=0,\omega)$. The singularity of the spinon branch appears at $\omega=-\pi J/2\sim -2\pi t^{2}/U=-0.785t$. The singularity comes from the flatness of the spinon dispersion near $k=0$. The lineshape is consistent with the DDMRG data in Fig. 1(a) except for the singularity of the holon branch. The singularity of the holon branch is recovered from the phase string effect~\cite{Suzuura}. The consistency indicates that Eq.~(\ref{effective}) is appropriate for the spectral weight in spin-charge separated systems.

Let us introduce the electron-phonon coupling, and take $g$ to be $0.5t$. Due to the spin-charge separation, each of the holons couples with phonons independently. Namely, $A_{h}(p,\varepsilon)$ is given by the spectra for a spinless carrier dressed with Einstein phonons. Figure 3(c) shows $A_{h}(p-\pi,\varepsilon)$ that splits into low-lying peaks and an incoherent part~\cite{Hohenadler,Sykora,Zhao,Zhang}. The split occurs at the anticrossing point $\varepsilon\sim 1.5t$ that is away from the top of the band ($p=\pi$) by $\omega_{0}$. At $\varepsilon\sim 1.5t$, we find a tiny spectral weight with a flat dispersion coming from the phonon branch. In Fig. 3(a), the split of one holon dispersion is illustrated. $A_{\rm eff}(k=0,\omega)$ is then given by a solid line in Fig. 3(b). A peak-dip-hump structure appears. The spectral weight lost by the dip is transfered to high energy region. The spinon branch, the dip and the broad holon branch originate from the low-lying peak, the anticrossing and the incoherent part of $A_{h}(p,\varepsilon)$, respectively. These features obtained by Eq.(\ref{effective}) are consistent with the DDMRG data, and thus the spin-charge separation is robust.

Next, we consider the origin of (iii) shown in Fig. 2(b) for $g\ge\omega_{0}$. As mentioned in the previous paragraph, the spinon branch can be expressed by the superposition of the low-lying peak in Fig. 3(c). It has been shown that the spectral weight of the low-lying peak decreases with increasing $g$~\cite{Hohenadler,Sykora,Zhao,Zhang}. Therefore, the weight of the spinon branch decreases with increasing $g$. 

\begin{figure}
\begin{center}
\includegraphics[width=13cm]{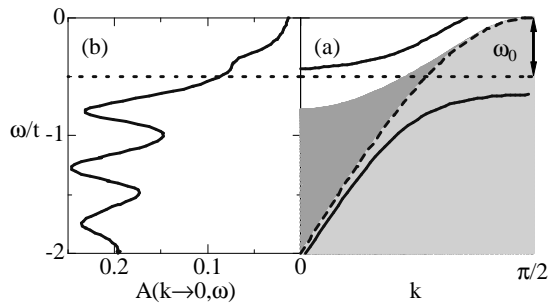}
\caption{(a) Shadded area: schematic view of the dispersion in a 1D Mott insulator; dashed (solid) line: one holon dispersion for $g=0$ ($g\ne 0$). A dotted line guides the eye. (b) $A(k\rightarrow 0,\omega)$ for $g=t$ and $\omega_{0}=0.5t$.}
\end{center}
\end{figure}

Finally, the origin of (iv) is considered. In Fig. 4, we see that the origin comes from the phonon branch. Even for the coupling $g=t>\omega_{0}$, the spectral weight of the phonon branch is very weak at $k\sim 0$~\cite{Hohenadler,Sykora,Zhao,Zhang}. This is the reason for the tiny weight seen in the DDMRG data. It is noted that the tiny weight is observed when a condition $\omega_{0}<\pi J/2$ is satisfied.

All of our interpretations for (i)-(iv) show robustness of the spin-charge separation. The robustness leads to the difference of the effect of phonon between 1D and 2D systems in the presence of electron correlation. In order to see the difference, we introduce a dimensionless parameter $\lambda=g^{2}/\omega_{0}W$ with the noninteracting band width $W$. There is a characteristic $\lambda$ value, $\lambda^{\ast}$~\cite{Kabanov}. For $\lambda<\lambda^{\ast}$, the lowest energy excitation is weakly dressed with phonons. As $\lambda$ approaches to $\lambda^{\ast}$, the excitation loses its weight rapidly. Then, the dominant low-energy excitation moves to a heavily dressed polaron. In the 1D Holstein model, $\lambda^{\ast}$ is estimated to be $\lambda^{\ast}\sim 1$~\cite{Bonca,Wellein}. In the 1D Hubbard-Holstein model, the spinon branch loses its weight for $\lambda^{\ast}\sim 1$ ($g^{\ast}\sim 1.4t$) as estimated from Fig. 2(b)~\cite{Capone}. This is because the spinon branch can be expressed by the superposition of the spectra for the low-energy excitation of the Holstein model. In 2D, on the other hand, $\lambda^{\ast}$ is close to $0.2$ in the $t$-$J$-Holstein model with $J=0.3t$, while $\lambda^{\ast}>0.6$ in the Holstein model~\cite{Mishchenko,Bauml,Prelovsek}. This means that the AF correlation helps the formation of the lattice polaron~\cite{Cappelluti,Wellein2,Ramsak}. As $J$ increases, the energy gain is caused by the AF correlation. Then, the coherent motion of a hole is suppressed. The suppression helps the formation of the lattice polaron.

Finally, let us discuss the ARPES data for a 1D Mott insulator SrCuO${}_{2}$ in the light of the DDMRG data. In this compound, high-energy ARPES experiments have been done, where the spinon and holon branches were observed~\cite{Kim}. Near the $\Gamma$ point, the intensity of the holon branch is smaller than that of the spinon branch. In addition, these branches do not exhibit singularities predicted by the Hubbard model~\cite{Sorella}. In Fig. 2(b), we show the DDMRG data for $\lambda=0.25$ ($g=0.707t$ and $\omega_{0}=0.5t$) that is appropriate for 1D cuprates~\cite{Gunnarsson}. The DDMRG data is consistent with the ARPES data except for the dip. The dip may be covered by the phonon dispersion. Finite temperature effects in the presence of $g$ might be the reason for the covering. 

In summary, we have examined the single-particle excitation spectrum in the 1D Hubbard-Holstein model at half-filling by performing the DDMRG calculations. We found the following four characteristic features: (i) peak-dip-hump structure, (ii) broad holon branch, (iii) the decrease of the spectral weight of the spinon branch, and (iv) the enhancement of the weight at low-binding energy side of the spinon branch. In the light of the spin-charge separation, these results are reproduced well by performing superposition of the spectra for a spinless carrier dressed with phonons. In 1D systems, the spin-charge separation is robust against the electron-phonon coupling, while in 2D systems, the AF spin correlation favors the formation of the lattice polaron. The ARPES data for SrCuO${}_{2}$ was consistent with our DDMRG data.

We thank P. Prelov\v{s}ek for discussions and for giving us a preprint prior to publication. This work was supported by NAREGI Nanoscience Project and Grant-in-Aid for Scientific Research from the Ministry of Education, Culture, Sports, Science and Technology of Japan, and CREST.

\end{document}